\documentclass[%
 aip,
 jmp,%
 amsmath,amssymb,
reprint,%
]{revtex4-1}

\usepackage{graphicx}
\usepackage{dcolumn}
\usepackage{bm}
\usepackage{epstopdf}
\usepackage{amssymb}
\usepackage{color}
\usepackage[colorlinks=true, letterpaper=true, pdfstartview=FitV, linkcolor=blue, citecolor=blue, urlcolor=blue]{hyperref}

\begin{document}

\title{Strain-tunable magnetic order and electronic structure in 2D CrAsS$_4$}
\author{Tengfei Hu}
\author{Wenhui Wan}
\author{Yanfeng Ge}
\author{Yong Liu}\email{yongliu@ysu.edu.cn, or ycliu@ysu.edu.cn}
\affiliation{State Key Laboratory of Metastable Materials Science and Technology Key Laboratory for Microstructural Material Physics of Hebei Province, School of Science, Yanshan University, Qinhuangdao 066004, China }

\begin{abstract}
  The effect of strain on the magnetic order and band structure of single-layer CrAsS$_4$ has been investigated by first-principles calculations based on density functional theory. We found that single-layer CrAsS$_4$ was an antiferromagnetic (AFM) semiconductor, and would have a phase transition from AFM state to ferromagnetic (FM) state by applying a uniaxial tensile strain of 2.99\% along the y-direction or compressive strain of 1.76\% along the x-direction. The underlying physical mechanism of strain-dependent magnetic stability was further elucidated as the result of the competition between the direct exchange and indirect superexchange interactions. Moreover, band gap exhibit a abrupt change along with phase transition of magnetic order. Our study provides an intuitional approach to design strain-modulated spintronic devices.
\end{abstract}


\maketitle


\maketitle

Since the initial boom of graphene research, other two-dimensional (2D) materials have been discovered, which possess highly tunable physical properties and immense potential for scalable device applications~\cite{1,2,3}. Among 2D materials, the monolayer structures of the transition-metal dichalcogenides (TMDCs) VX$_2$ (X=S, Se)~\cite{4}, trihalides CrX$_3$ (X=F, Cl, Br, I)~\cite{5}, ternary chalcogenides CrXTe$_3$ (X=Si, Ge, Sn)~\cite{6,7}, and transition-metal phosphorus (P) trichalcogenides MPX$_3$ (M=Fe, Mn, Ni; X=S, Se)~\cite{8,9,10}, exhibit interesting magnetic properties. The controllable spins of such 2D materials can be used to process information, replace current hard disks and even develop the quantum computation. Further, electrostatic shielding effect becomes weaker along the z-axis than that in the bulk counterparts of these materials owing to the dimensional reduction; thus, field-effect devices can be constructed easily for use as sensors or logic devices.

Control of the magnetic order is an important issue for spintronic-device applications. Strain engineering has been universally adopted to tune properties of nanomaterials using substrate lattice mismatching~\cite{11,12}. Additionally, it has been demonstrated that external strain can tune the bandgap of many 2D materials, such as single-layer  MoS$_2$~\cite{11}. Similarly, the magnetic order of the transition-metal P trichalcogenide MnPSe$_3$ has been adjusted using biaxial tensile strain, causing a magnetic phase transition from AFM to FM~\cite{13}. This transition has been determined in many other single-layer van der Waals (vdW) magnetic materials, such as the trihalides CrX$_3$ (X=F, Cl, Br, I), as well~\cite{14}.

In this paper, first-principle calculations are used to investigate the electronic and magnetic properties of single-layer CrAsS$_4$.  It is determined that a magnetic phase transition from AFM to FM occurs when the layer is stretched by 2.99\% along the y-direction or compressed by 1.76\% along the x-direction. By investigating the physical mechanism underlying this phenomenon, it is found that the magnetic order results from competition between direct exchange  and indirect superexchange interactions. When the layer is compressed along the x-direction, the bandgap decreases and undergoes a transition from a direct to an indirect bandgap; conversely, stretching it along the y-direction also causes the bandgap to decrease, but it remains a direct bandgap.

Kohn-Sham density-functional theory (DFT) calculations are performed using the projector augmented wave (PAW) method, as implemented in the plane-wave code VASP~\cite{15,16,17}. The wave functions of the valence electrons (3d$^{5}$4s$^{1}$ for Cr, 3s$^{2}$3p$^{4}$ for S, 4s$^{2}$4p$^{3}$ for As) were expanded in plane-wave basis sets, with a kinetic energy cutoff at 500 eV. We used the Perdew-Burke-Ernzerhof (PBE) type generalized gradient approximation (GGA) in the exchange-correlation functional~\cite{18}. The k-points in the Brillouin zone were sampled using the Monkhorst-Pack scheme~\cite{19}, employing $7\times9\times9$ and $17\times19\times1$ meshes to calculate bulk and single-layer Cr AsS$_4$, respectively. A conjugate-gradient algorithm was employed for geometry optimization using convergence criteria of 10$^{-7}$ eV for the total energy, and 0.01 eV/{\AA} for Hellmann-Feynman force components. We used GGA+U to treat the strong on-site Coulomb interaction~\cite{20}, adopting Hubbard U term and J term of 3 eV and 1.4 eV for Cr, respectively. We included the vdW correction in all the calculations using Grimme's method (DFT-D2)~\cite{21}. We inserted a 20 $\AA$  vacuum slab into the calculation grid to avoid the interactions between periodic images.

\begin{figure}[htb]
  \centering
  \includegraphics[width=.48\textwidth]{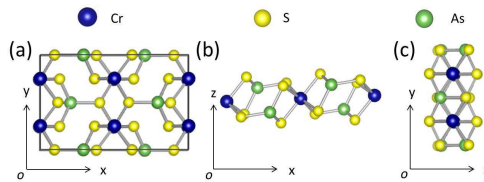}
 \caption{(Color online) The crystal structure of single-layer CrAsS$_4$ as seen from the (a) z-direction, (b) y-direction and (c) x-direction.}\label{fig1}
\end{figure}


Fig.~\ref{fig1} illustrates the unit cell of single-layer CrAsS$_4$. Each unit cell includes four Cr$^{3+}$ ions and its chemical formula can be written as Cr$^{3+}$[AsS$_4$]$^{3-}$. Cr atoms are sandwiched between S and As atoms, and an As atom connects four S atoms in [AsS$_4$]$^{3-}$. According to the Hund's rule, Cr atoms with 3d electron have the completely unoccupied e$_g$ electronic orbits and fully occupied t$_{2g}$ electronic orbits. The electrons of fully occupied levels have the same spin direction.

\begin{table}[h]
\newcommand{\PreserveBackslash}[1]{\let\temp=\\#1\let\\=\temp}
   \newcolumntype{C}[1]{>{\PreserveBackslash\centering}p{#1}}
   \newcolumntype{L}[1]{>{\PreserveBackslash\raggedright}p{#1}}
   \newcommand{\tabincell}[2]{\begin{tabular}{@{}#1@{}}#2\end{tabular}}
\begin{ruledtabular}
\caption{Lattice parameters a, b, and electron magnetic moment m for per Cr atom.}\label{table1}
\begin{tabular}{lccccc}
   order & a(\AA) & b(\AA) & m($\mu_B$)\\
   \hline
   FM & 11.40 & 7.25 & 3.017\\
   AFM1 & 11.43 & 7.24 & 2.998\\
   AFM2 & 11.45 & 7.19 & 2.935\\
   AFM3 & 11.45 & 7.18 & 2.939\\
\end{tabular}
\end{ruledtabular}
\end{table}

Table \ref{table1} lists the optimized lattice parameters for different magnetic configurations. Different magnetic configurations are determined to have only slightly different lattice constants and magnetic moments of the Cr atoms in the different structures are all nearly 3 $\mu_B$. This result is similar to the previous reported about CrPS$_4$~\cite{22}.

\begin{figure}[htb]
  \centering
  \includegraphics[width=.48\textwidth]{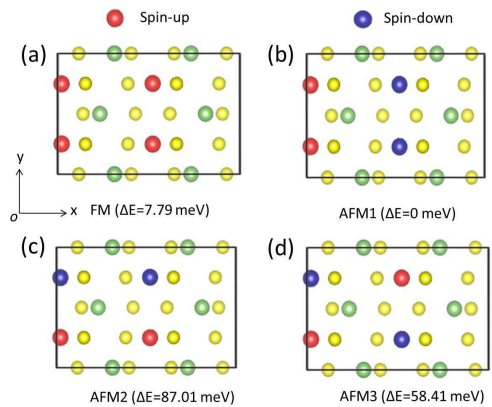}
 \caption{(Color online) Four possible magnetic configurations for single-layer CrAsS$_4$, with their relative energy differences $\Delta$E (meV/unit cell) with respect to the AFM1 configuration.}\label{fig2}
\end{figure}

To determine the ground-state magnetic order of single-layer CrAsS$_4$, we selected one FM and three possible AFM magnetic orders, as shown in Fig.~\ref{fig2}, of which AFM1 is the most stable. The energy differences $\Delta E$ relative to AFM1 configurations are 7.79, 87.01, and 58.41 meV for the FM, AFM2, and AFM3 orders, respectively. Because the non-magnetic state has a much higher energy than the magnetic states, we neglect it in the following discussion, and the following AFM calculations use the AFM1 phase.

\begin{table}[h]
\begin{ruledtabular}
\caption{Elastic constants C$_{11}$, C$_{12}$ and C$_{22}$ (N/m). Poisson's ratios $\nu_1$ and $\nu_2$ of single-layer CrAsS$_4$ in AFM1 state.}\label{table2}
\begin{tabular}{lccccc}
   C$_{11}$ & C$_{12}$ & C$_{22}$ & $\nu_1$ &  $\nu_2$\\
   \hline
   64.29 & 20.56 & 60.58 & 0.32 &  0.34\\
\end{tabular}
\end{ruledtabular}
\end{table}

Table~\ref{table2} lists the elastic constants and Poisson's ratios for AFM1. The elastic constants clearly satisfy Born¡¯s stability criterion for single-layer materials~\cite{23}, i.e., C$_{11}{>}$0, C$_{22}{>}$0 and C$_{11}$-C$_{12}{>}$0, indicating that they are mechanically stable.

Defining Poisson's ratios as ${\nu _1} = \frac{{{C_{12}}}}{{{C_{11}}}}$, ${\nu _2} = \frac{{{C_{12}}}}{{{C_{22}}}}$, we found $\nu_1$ $<$ $\nu_2$, which indicates that the stiffness along the Cr chain is less than that in the direction perpendicular to the Cr chain. Furthermore, we evaluate the stability of single-layer CrAsS$_4$ according to the formation energy $\emph{E}_f$~\cite{24},

\begin{equation}
{E_f}{\rm{ = }}\frac{{{E_{2D}}}}{{{N_{2D}}}} - \frac{{{E_{3D}}}}{{{N_{3D}}}},
\end{equation}
where $\emph{E}_{2D}$ and $\emph{E}_{3D}$ are total energy of single-layer and bulk structures, respectively. The quantities $\emph{N}_{2D}$ and $\emph{N}_{3D}$ denote the number of atoms in the corresponding unit cells. We found that $\emph{E}_f$  of 39.61 meV/atom, is smaller and larger than those obtained in previous research on actually synthesized VS$_2$ (90 meV/atom)~\cite{25} and on CrPS$_4$ (1.44 meV/atom)~\cite{22}, which make it quite possible to mechanically exfoliate single-layer CrAsS$_4$ from the bulk.

\begin{figure}[htb]
\centering
\includegraphics[width=.48\textwidth]{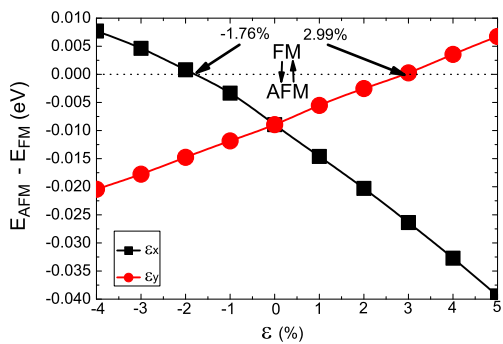}
\caption{(Color online)Possible paths for electrons hopping in magnetic-exchange interactions (a) between nearest-neighbor and (b) between second-nearest-neighbor in single-layer CrAsS$_4$. (Anti-)parallel arrows represents the (anti-)ferromagnetic exchange interactions.}\label{fig3}
\end{figure}

Fig.~\ref{fig3} shows the energy difference between the magnetic ground states of the two magnetic orderings under different strains. Herein, the strain is defined as $\varepsilon  = \frac{{(a - {a_0})}}{{{a_0}}}$, where $\emph{a}_0$ and $\emph{a}$ is the lattice constant of unstrained and strained single-layer CrAsS$_4$, respectively. When a uniaxial tensile strain of 2.99\% along the y-direction, the stability of the FM phase increases. It is determined that an AFM-FM phase transition also occurs when a uniaxial compressive strain of 1.76\% is applied along the x-direction. Such adjustments of the spin order can be applied in spintronics devices.

The transition of single-layer CrAsS$_4$ from AFM-FM can be understood in terms of exchange interaction. The effective Heisenberg Hamiltonian is used to characterize the magnetic properties of CrAsS$_4$,
\begin{equation}
H =  - \sum\limits_{ < ij > } {{J_{ij}}{S_i}{S_j}},
\end{equation}
where $\emph{J}_{ij}$ represents the exchange interactions of over all neighbor Cr-Cr pairs. Herein $\emph{S}_i$ represents the spin of atom $\emph{i}$, and we only consider two nearest neighbors located along x and y directions. To compute the $\emph{J}$, we write the energy of a single-layer CrAsS$_4$ unit cell in the FM, AFM1 and AFM3 configurations in the following forms:

\begin{equation}
{E_{AFM1}} = {E_0} + 4{J_x}{S^2} - 4{J_y}{S^2},
\end{equation}
\begin{equation}
{E_{FM}} = {E_0} - 4{J_x}{S^2} - 4{J_y}{S^2},
\end{equation}
\begin{equation}{E_{AFM3}} = {E_0} + 4{J_x}{S^2} + 4{J_y}{S^2}.
\end{equation}
The quantity $\emph{E}_0$ is the energy of single-layer CrAsS$_4$ without spin polarizations~\cite{26} and $\emph{S}$=3/2. A positive $\emph{J}$ value represents an FM state and a negative one represents an AFM state. With $\emph{J}_y$= 3.03 meV, $\emph{J}_x$ = -0.49 meV, the ground state is AFM1.

\begin{figure}[htb]
  \centering
  \includegraphics[width=.48\textwidth]{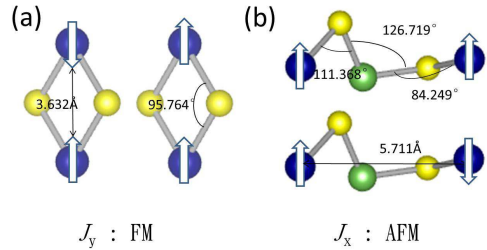}
 \caption{Possible paths for electrons hopping of (a) the nearest neighbor and (b) the second nearest neighbor magnetic exchange intercations in single-layer CrAsS$_4$. (anti-)parallel arrangement arrow represents the (anti-)ferromagnetic exchange interactions.}\label{fig4}
\end{figure}

Further, the microscopic origin of the exchange interactions in single-layer CrAsS$_4$ is considered. As shown in Fig.~\ref{fig4} (a), the contribution $\emph{J}_y$ is obtained from two kinds of exchange interactions: (1) direct exchange interactions among Cr-Cr, because Cr$^{3+}$ has a t$_{2g}^3$e$_g^0$ configuration that is AFM in character, and (2) superexchange interactions among Cr-S-Cr with an anionic mediation S. The Cr-S-Cr bond angle (95.764$^\circ$) is nearly 90$^\circ$, and according to the Goodenough-Kanamori-Anderson (GKA) rules~\cite{27,28}: if a cation-anion-cation bond angle is nearly 90$^\circ$, the system prefers FM order; conversely it is AFM for 180$^\circ$. The Cr-S-Cr path is preferential for FM order, because the indirect FM superchange interaction is dominant, $\emph{J}_y$ exhibits FM properties, which is consistent with the previous research on CrSiTe$_3$~\cite{29}.

\begin{figure}[htp!]
  \centering
  \includegraphics[width=.48\textwidth]{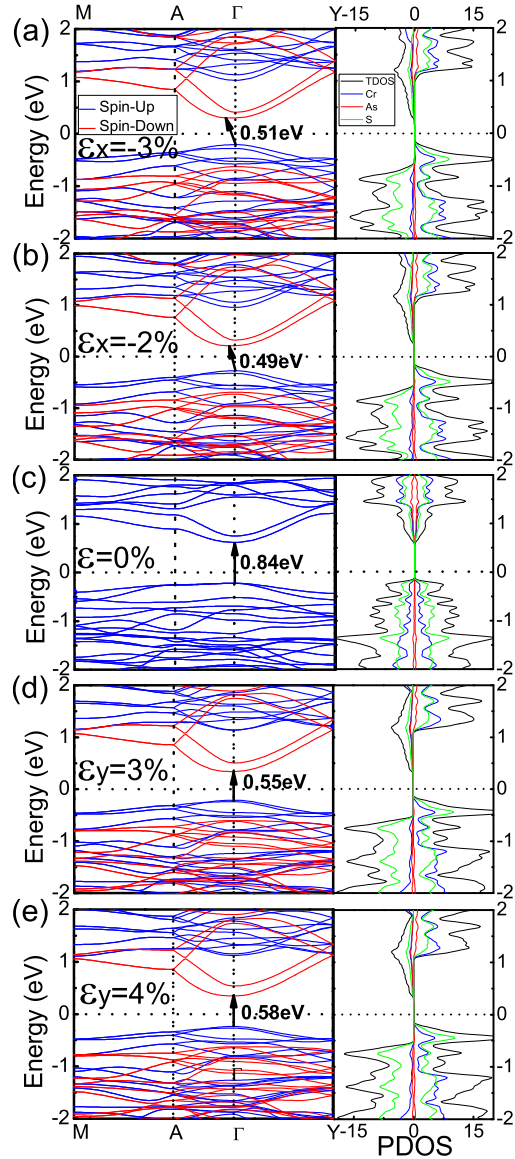}
 \caption{(Color online)Band structures and partial densities of states (PDOS) for single-layer CrAsS$_4$ under uniaxial strain. Panels (a) and (b) correspond to compressive strain of 3\% and 2\% along the x-direction, respectively. Panels (d) and (e) correspond to tensile strain of 3\% and 4\% along the y-direction, respectively. Panels (c) are band structure and PDOS of the unstrained layer, respectively.}\label{fig5}
\end{figure}

As shown in Fig.~\ref{fig4}(b), a long-range Cr-S-As-S-Cr super-superexchange interaction occurs along the x-direction. This is similar to cation-anion-cation superexchange interaction, thus we can also use the 90$^\circ$ and 180$^\circ$ criterion herein. For an intermediate angle between 90$^\circ$ and 180$^\circ$, a crossover angle from FM to AFM should exist, thus it is concluded that the crossover angle of 127$\pm$0.6$^\circ$ is related to the FM-AFM transition~\cite{30}. It is determined that the angles along the x-direction are smaller than 127$^\circ$, but the path of electron hopping is long-range; thus, the direct exchange interaction of Cr-Cr is stronger than  indirect super-superexchange interaction. Thus, the term $\emph{J}_x$ exhibits AFM properties, and the ground state shows AFM characteristics.

\begin{figure}[htb]
    \centering
  \includegraphics[width=.48\textwidth]{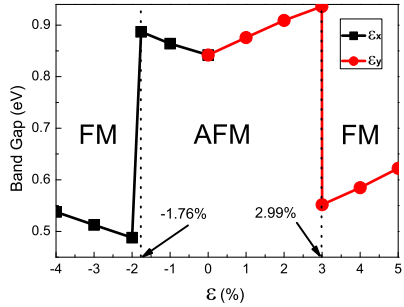}
    \caption{(Color online) Variation of the band gap at different uniaxial strain.}\label{fig6}
  \end{figure}

The short-range direct exchange interaction is sensitive to changes in the Cr-Cr distance, whereas the long-range indirect superexchange interaction is relatively insensitive to such changes. When a uniaxial tensile strain of 3\% is applied along the y-direction, the Cr-Cr distance increases and the short-range direct exchange interaction rapidly decreases. However, the long-range indirect superexchange interaction decreases relatively slowly; thus it causes $\emph{J}_y$ to prefer FM order. Although the distance between the two Cr atoms along the x-direction changes a little, the super-superexchange interaction increases, i.e., the bond angle Cr-S-P, S-P-S and P-S-Cr change from 111.368$^\circ$, 126.719$^\circ$ and 84.249$^\circ$ to 111.561$^\circ$, 126.094$^\circ$ and 84.674$^\circ$. This causes $\emph{J}_x$ becomes positive by 0.02 meV, the system presents FM order.

When a uniaxial compressive strain of 2\% is applied along the x-direction, the changes in the Cr-Cr distance and the superexchange interaction change are not obvious; thus $\emph{J}_y$ also prefers FM order. However, for the two Cr atoms along the x-direction, the super-superexchange interaction increases more than the direct exchange interaction; thus, $\emph{J}_x$ becomes positive by 0.05 meV, and the system presents FM order.

As shown in Fig.~\ref{fig5}(c), single-layer CrAsS$_4$ is a direct band gap semiconductor, with the band gap value of 0.84 eV. The valence-band maximum (VBM) contribution mainly includes the S and Cr atoms, whereas the conduction-band minimum (CBM) contribution includes all three atoms. According to Fig.~\ref{fig5}(a) and (b), 3\% and 2\% compressions cause the bandgap to decrease from 0.84 eV to 0.51 eV and 0.49 eV, respectively. Moreover, for a 2\% compression along the x-direction, the CBM shifts slightly, and the material becomes an indirect-bandgap semiconductor. Additionally, it is determined that applying a 3\% compression causes a larger shift; however, according to Fig.~\ref{fig5}(d) and (e), it is still a direct-bandgap semiconductor; the VBM and CBM are closest at the $\Gamma$ point, and the bandgap decreases to 0.55 eV and 0.58 eV owing to crystal-field splitting. Comparison between Fig.~\ref{fig5}(e) with Fig.~\ref{fig5}(c) shows the S and Cr atoms of spin-up near the VBM has a higher peak and more similar picture; thus, the degrees of electron localization and hybridization increase. This indicates that the indirect exchange interaction plays an important role in the FM coupling. It is determined that all the PDOS under the strain correspond to spin-down, with main contribution from the S atom and slight contributions from other atoms.

Figure~\ref{fig6} summaries the variation of the bandgap under different uniaxial strains. Whether the strain is compressive or stretching, it is determined that the bandgap decreases considerably in a magnetic phase transition and as we continue to stretch or compress, the bandgap increases.

In summary, the magnetic and electronic properties of single-layer CrAsS$_4$ via DFT have been studied. Single-layer CrAsS$_4$ is an AFM semiconductor. The calculations of mechanical properties and formation energy ensure the stability and the possibility of preparation of single-layer CrAsS$_4$. Application of a uniaxial tensile strain of 2.99\% along the y-direction or a compressive strain of 1.76\% along the x-direction causes a magnetic phase transition from AFM to FM, and the bandgap also decreases. The physical mechanism underlying this strain-tunable magnetic order is the competition between direct exchange and indirect superexchange interactions. Thus, our study shows that uniaxial strain can modulate the magnetic order of this material, it is a suitable candidate for spintronic devices.


This work was supported by the Specialized Research Fund for the Doctoral Program of Higher Education of China (Grant No.2018M631760), the Project of Heibei Educational Department, China(No.ZD2018015 and QN2018012), the Advanced Postdoctoral Programs of Hebei Province (No.B2017003004) and the Natural Science Foundation of Hebei Province (No. A2019203507).



\nocite{*}
\bibliography{hu}
\end{document}